\documentclass[preprint]{JHEP3} % 10pt is ignored!

\JHEPspecialurl{http://jhep.sissa.it/JOURNAL/JHEP3.tar.gz}

\usepackage{epsfig,multicol,bbm}
\usepackage{latexsym} 
\usepackage{amssymb}  

\newcommand\fverb{\setbox\pippobox=\hbox\bgroup\verb}
\newcommand\fverbdo{\egroup\medskip\noindent%
			\fbox{\unhbox\pippobox}\ }
\newcommand\fverbit{\egroup\item[\fbox{\unhbox\pippobox}]}
\newbox\pippobox
\newcommand{\e}{ {\rm e} }

\newcommand{\tr}{\hbox{tr}}
\newcommand{\Tr}{\hbox{Tr}}
\newcommand{\Gammatpi}{\Gamma_{\mbox{\scriptsize 2PI}}}

\newcommand{\beq}{\begin{equation}}
\newcommand{\eeq}{\end{equation}}
\newcommand{\ba}{\begin{array}}
\newcommand{\bea}{\begin{eqnarray}}
\newcommand{\ea}{\end{array}}
\newcommand{\eea}{\end{eqnarray}}

\newcommand{\nn}{\nonumber \\}
\newcommand\eqn[1]{(\ref{#1})}      % parentheses around the LaTex "ref" macro
\newcommand\Eqn[1]{Eq.~(\ref{#1})}  % includes ``Eq.'' in front

\def\slashchar#1{\setbox0=\hbox{$#1$}           % set a box for #1 
   \dimen0=\wd0                                 % and get its size
   \setbox1=\hbox{/} \dimen1=\wd1               % get size of /
   \ifdim\dimen0>\dimen1                        % #1 is bigger
      \rlap{\hbox to \dimen0{\hfil/\hfil}}      % so center / in box
      #1                                        % and print #1
   \else                                        % / is bigger
      \rlap{\hbox to \dimen1{\hfil$#1$\hfil}}   % so center #1
      /                                         % and print /
   \fi}

\title{2PI effective action for gauge theories: Renormalization}

\author{Urko Reinosa\\
	Institut f{\"u}r Theoretische Physik, Universit{\"a}t Heidelberg,\\
	Philosophenweg 16, 69120 Heidelberg, Germany.\\
	E-mail: \email{u.reinosa@thphys.uni-heidelberg.de}}
\author{Julien Serreau\thanks{Present address: Laboratoire de Physique Th\'eorique, B\^atiment 210,
	Universit\'e Paris 11 - Sud, 91405 Orsay Cedex, France. LPT is unit\'e mixte de recherche UMR 8627 (CNRS, Universit\'e Paris 11).}\\
	Astro-Particule et Cosmologie,\thanks{APC is unit\'e mixte de recherche UMR 7164 (CNRS, Universit\'e Paris 7, CEA, Observatoire de Paris).} Universit\'e Paris 7 - Denis Diderot,\\
	11, place Marcelin Berthelot 75231 Paris Cedex 05, France\\
	E-mail: \email{julien.serreau@th.u-psud.fr}}
\preprint{APC-06-19\\HD-THEP-06-6\\LPT-06-30}	% OR: \preprint{Aaaa/Mm/Yy\\Aaa-aa/Nnnnnn}
\abstract{
We discuss the application of two-particle-irreducible (2PI) functional techniques to gauge theories, focusing on the issue of non-perturbative renormalization. In particular, we show how to renormalize the photon and fermion propagators of QED obtained from a systematic loop expansion of the 2PI effective action. At any finite order, this implies introducing new counterterms as compared to the usual ones in perturbation theory. We show that these new counterterms are consistent with the 2PI Ward identities and are systematically of higher order than the approximation order, which guarantees the convergence of the approximation scheme. Our analysis can be applied to any theory with linearly realized gauge symmetry. This is for instance the case of QCD quantized in the background field gauge. 
}


\begin{document} 

%\maketitle  IS IGNORED %%%%%%%%%%%

\section{Introduction}
\label{sec:intro}

Functional techniques based on the two-particle-irreducible (2PI) effective action~\cite{Luttinger:1960ua,Cornwall:1974vz} in quantum field theory have attracted a great deal of interest in recent years (for reviews, see e.g.~\cite{Blaizot:2003tw,Berges:2003pc}). They provide a powerful tool to devise systematic non-perturbative approximations, of particular interest in numerous physical situations where standard expansion schemes are badly convergent. Timely examples include scalar or gauge field theories at high temperature~\cite{Blaizot:1999ip,Berges:2004hn,Blaizot:2005wr,Aarts:2003bk}, or far-from-equilibrium dynamics~\cite{Berges:2000ur,Aarts:2002dj,Berges:2002cz,Berges:2002wr}.

The systematic implementation of 2PI techniques for gauge theories has been postponed for a long time due to formal intricacies, see e.g.~\cite{Arrizabalaga:2002hn,Mottola:2003vx}.\footnote{For discussion of higher effective actions in the context of gauge theories, see e.g.~\cite{Berges:2004pu}.} One central problem is that of gauge invariance and the possibility or not to define gauge independent quantities within a given approximation of the 2PI effective action. Although no procedure exists to construct such gauge independent quantities, general results~\cite{Arrizabalaga:2002hn} indicate that gauge dependent contributions are parametrically suppressed in powers of the coupling (see also~\cite{Andersen:2004re}). Existing work is however based on considerations at the level of the bare, regularized theory and cannot be extrapolated to the continuum limit prior to an understanding of the issue of renormalization in this context. The latter is also of obvious importance for the purpose of performing practical calculations within the 2PI formalism (see e.g.~\cite{VanHees:2001pf,Berges:2004hn}).

The elimination of ultra-violet divergences in the 2PI framework has been understood only recently in theories with scalar~\cite{vanHees:2001ik,Berges:2005hc} and fermionic~\cite{Reinosa:2005pj} degrees of freedom. This is actually possible thanks to the intrinsic self-consistency of the 2PI approach.\footnote{2PI renormalization has also recently been discussed in relation with functional renormalization group techniques in \cite{Pawlowski:2005xe}.} Similar to standard perturbation theory, 2PI renormalization requires one to consistently include, at each order of approximation in a systematic (e.g. loop or $1/N$) expansion, all local counterterms of mass dimension lower or equal to four, consistent with the symmetries of the theory~\cite{Berges:2005hc}. The essential difference in the 2PI framework is that, since the theory is parametrized in terms of both the one- and two-point functions, there are new possible counterterms at each finite order.\footnote{Of course, in the exact theory, these new counterterms coincide with the usual ones: there are no new parameters in the theory.} These counterterms have non-perturbative expressions since they absorb the UV divergences of infinite series of perturbative diagrams. 

As mentioned previously, counterterms are restricted by the symmetries of the theory: they must be consistent with the 2PI Ward identities. The cases studied so far concern theories with linear global symmetries (including spontaneously broken symmetries), for which the 2PI Ward identities are easily solved~\cite{Berges:2005hc}. In this paper, we address the question of 2PI renormalization for QED. Our analysis is directly applicable to any theory with linearly realized gauge symmetry. We concentrate on the renormalization of the physical photon and fermion propagators, obtained as saddle points of the 2PI effective action, which play a central role in this formalism. The analysis of the 2PI-resummed effective action, which encodes all the vertices of the theory, will be presented in a subsequent paper~\cite{JU2}.

Again, in the case of linearly realized gauge symmetry, the 2PI Ward identities are easily solved. We show that, besides the analog of the standard field strength, mass and charge counterterms, they allow for new counterterms, which have no analog in perturbation theory. The need for the latter is related to the fact that, in contrast to the standard 1PI Ward identities, the 2PI Ward identities do not prevent the occurrence of non-transverse divergent contributions to the two- and four-photon functions at any finite approximation order. We show that these divergences can indeed be absorbed in the above mentioned extra counterterms. We also show that the latter are systematically of higher order than the approximation order in a 2PI loop-expansion, provided one imposes suitable renormalization conditions. This is a crucial point since it guarantees that the employed approximation scheme converges towards the correct exact theory.

In Sec.~\ref{sec:generalities}, we introduce relevant definitions and discuss the symmetry properties of the 2PI effective action for QED in the linear covariant gauge. Since our main purpose is renormalization, we focus on the vacuum theory. In Sec.~\ref{sec:counterterms}, we classify the counterterms consistent with these symmetries at two-loop order. Section~\ref{sec:renormalization} proceeds with the renormalization of the fermion and photon propagators, which involves the renormalization of the four-photon vertex. The generalization of the renormalization program to higher-loop orders is discussed in Sec.~\ref{sec:higher}.

%%%%%%%%%%%%%%%%%%%%   2PI effective action for QED
\section{2PI effective action for QED}
\label{sec:generalities}

%%%%%%%%%%%%%%%   Generalities
\subsection{Generalities}
\begin{figure}[tbp]
\begin{center}
\includegraphics[width=7.5cm]{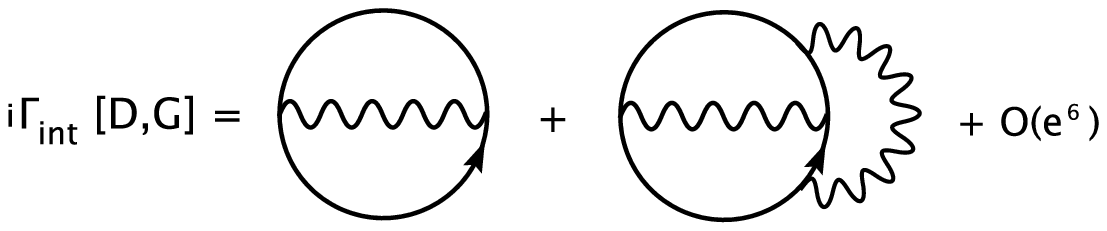}
\caption{\label{fig:loopQED}Leading contributions to $\Gamma_{\rm int}[D,G]$ in a loop expansion.}\label{fig:phi}
\end{center}
\end{figure}

We consider QED in the covariant gauge and use dimensional regularization. The gauge-fixed classical action reads 
\begin{equation}
\label{eq:classact}
S=\int_x \left\{\bar\psi\Big[i\slashchar{\partial}-e\slashchar{A}-m\Big]\psi+\frac{1}{2}A_\mu\Big[g^{\mu\nu}\partial^2-(1-\lambda)\partial^\mu\partial^{\nu}\Big]A_\nu\right\}\,,
\end{equation}
where $\int_x\equiv\int d^dx$ and $\lambda$ is the gauge-fixing parameter. Aside from the gauge-fixing term, the classical action is invariant under the gauge transformation
\begin{equation}\label{eq:gauge}
\psi(x)\rightarrow e^{i\alpha(x)}\,\psi(x)\,,\quad
\bar\psi(x)\rightarrow e^{-i\alpha(x)}\,\bar\psi(x)\,,\quad
A_\mu(x)\rightarrow A_\mu(x)-\frac{1}{e}\,\partial_\mu\alpha(x)\,,\nonumber\\
\end{equation}
where $\alpha(x)$ is an arbitrary real function. In this paper, we consider the 2PI effective action~\cite{Cornwall:1974vz} at vanishing fields,\footnote{Renormalization of the complete effective action with non-vanishing fields will be described elsewhere \cite{JU2}.} which is a functional of the full fermion and photon propagators $D_{\alpha\beta}(x,y)$ and $G_{\mu\nu}(x,y)$:
\begin{equation}\label{eq:2PI}
\Gamma_{\rm 2PI}[D,G]=-i\Tr\ln D^{-1}-i\Tr\, D_0^{-1}D+\frac{i}{2}\Tr\ln G^{-1}+\frac{i}{2}\Tr\,G_0^{-1}G+\Gamma_{\rm int}[D,G]\,,
\end{equation}
where the trace {\rm Tr} includes integrals in configuration space and sums over Dirac and Lorentz indices. The free inverse propagators are given by
\begin{eqnarray}
\label{eq:freepf}
iD_{0,\alpha\beta}^{-1}(x,y) & = & \left[\,i\slashchar{\partial}_x-m\,\right]_{\alpha\beta}\delta^{(4)}(x-y)\,,\\
\label{eq:freepb}
iG_{0,\mu\nu}^{-1}(x,y) & = & \left[\,g_{\mu\nu}\partial_x^2-
(1-\lambda)\partial^x_\mu\partial^x_\nu\,\right]\delta^{(4)}(x-y)\,.
\end{eqnarray}
The functional $\Gamma_{\rm int}[D,G]$ is the infinite series of closed two-particle-irreducible (2PI) diagrams with lines corresponding to $D$ and $G$ and with the usual QED vertex. Fig.~\ref{fig:phi} shows the first diagrams contributing to it in a loop expansion.

The physical propagators $\bar D$ and $\bar G$ can be obtained from the condition that the 2PI
functional be stationary in propagator space, which can be written as
\bea
\label{eq:motion1}
 \bar D_{\,\alpha\beta}^{-1}(x,y)-D^{-1}_{0,\alpha\beta}(x,y)=
 i\left.\frac{\delta\Gamma_{\rm int}[D,G]}{\delta D_{\beta\alpha}(y,x)}\right|_{\bar D,\bar G}
 &\equiv&-\bar\Sigma_{\alpha\beta}(x,y)\,,\\
\label{eq:motion2}
 \bar G_{\mu\nu}^{-1}(x,y)- G_{0,\mu\nu}^{-1}(x,y)=
 -2i\left.\frac{\delta\Gamma_{\rm int}[D,G]}{\delta G^{\nu\mu}(y,x)}\right|_{\bar D,\bar G}
 &\equiv&-\bar{\Pi}_{\mu\nu}(x,y)\,.
\eea
Although these equations are exact as they stand, it is usually not possible to compute $\Gamma_{\rm int}[D,G]$ completely and one has to rely on approximations. For instance, standard perturbation theory corresponds to a systematic expansion of Eqs.~(\ref{eq:motion1})-(\ref{eq:motion2}) in powers of the free propagators $D_0$ and $G_0$. As mentioned previously, this turns out to be badly convergent in a number of cases, including thermal equilibrium at high temperatures~\cite{Blaizot:2003tw}, or non-equilibrium situations \cite{Berges:2003pc}. In contrast, the 2PI resummation scheme consists in constructing systematic non-perturbative approximations -- e.g. using a loop, or $1/N$ expansion -- of the functional $\Gamma_{\rm int}[D,G]$ and solving the resulting non-linear equations (\ref{eq:motion1})-(\ref{eq:motion2}) without further approximations\footnote{This last point is crucial. It ensures that the conservation laws corresponding to global symmetries are respected at any approximation order. It also guarantees that the approximation scheme is non-perturbatively renormalizable~\cite{vanHees:2001ik,Berges:2005hc,Reinosa:2005pj}.}: each new contribution to $\Gamma_{\rm int}[D,G]$ gives rise to the resummation of a new infinite class of perturbative diagrams. In this paper, we shall concentrate on the 2PI loop-expansion of $\Gamma_{\rm int}[D,G]$. It is important to stress at this point, that the non-perturbative propagators obtained from $\Gamma_{\rm int}[D,G]$ at $L$-loop order agree with the perturbative expansion of the exact propagators up to, and including, order $e^{2(L-1)}$. This is because the neglected higher-loop contributions to $\Gamma_{\rm int}[D,G]$ contribute to $\bar D$ and $\bar G$ only at order $e^{2L}$.

%%%%%   Symmetries
\subsection{Symmetries}
The gauge transformation (\ref{eq:gauge}) for the connected two-point functions $D$ and $G$ reads
\begin{equation}\label{eq:gauge_prop}
 D(x,y)\to D^\alpha(x,y)\equiv e^{i\alpha(x)}D(x,y)e^{-i\alpha(y)}\,\,,
 \,\, G(x,y)\to G^\alpha(x,y)\equiv G(x,y)\,.
\end{equation}
Any diagram in $\Gamma_{\rm int}[D,G]$ only contains closed fermion loops and is then trivially invariant under such a transformation. It follows that the functional $\Gamma_{\rm int}[D,G]$ satisfies the 2PI Ward identity (see also \cite{JU2})\footnote{Notice that this does not depend on the regularization procedure.}
\beq
\label{eq:gaugeinv}
 \Gamma_{\rm int}[D,G]=\Gamma_{\rm int}[D^\alpha,G^\alpha]\,,
\eeq
at any loop order. It is important to realize that the gauge symmetry (\ref{eq:gaugeinv}) of the 2PI functional does not impose any constraint on the propagators $\bar D$ and $\bar G$, defined by Eqs.~(\ref{eq:motion1})-(\ref{eq:motion2}). This is to be contrasted with the usual Ward identities of the 1PI effective action, which impose strong constraints on $n$-point vertices, obtained as functional derivatives of the latter. In particular, the photon polarization tensor is constrained to be transverse in momentum space. In the present case, the propagators $\bar D$ and $\bar G$ are obtained, not as functional derivatives of an action, but as solutions of a stationarity condition for the 2PI functional and the resulting symmetry constraints are very different. Although the functional $\Gamma_{\rm int}[D,G]$ is gauge invariant at any loop order, there is no reason for the corresponding photon polarization tensor in momentum space $\bar \Pi_{\mu\nu}(k)$, as obtained from Eq.~(\ref{eq:motion2}), to satisfy the exact Ward identity $k^\mu\,\bar\Pi_{\mu\nu}(k)=0$. Since at $L$-loop order, $\bar G$ agrees with the perturbative expansion of the exact photon propagator up to, and including, order $e^{2(L-1)}$, one has instead $k^\mu\,\bar\Pi_{\mu\nu}(k)=\mathcal{O}(e^{2L})$. Only in the exact theory, where one takes into account all  contributions to $\Gamma_{\rm int}[D,G]$ (i.e. $L\rightarrow\infty$), does $\bar\Pi_{\mu\nu}(k)$ becomes exactly transverse.

\vspace{.2cm}
Equations (\ref{eq:motion1})-(\ref{eq:motion2}) are divergent and require renormalization. The purpose of the present paper is to show that the latter can be achieved at any order in the 2PI loop-expansion, in a way which is consistent with the gauge symmetry (\ref{eq:gaugeinv}).

%%%%%%%%%%%%%%%%%%%%   Counterterms
\section{Counterterms}
\label{sec:counterterms}
Non-perturbative renormalization of self-consistent equations such as (\ref{eq:motion1})-(\ref{eq:motion2}) has been analyzed in details in recent years in theories with scalar~\cite{vanHees:2001ik,Berges:2005hc} and fermionic~\cite{Reinosa:2005pj} fields. The result of such an analysis is that, in order to obtain finite results in a systematic 2PI loop-expansion, one has to include all possible local counterterms of mass dimension $d_M\le 4$ consistent with the symmetries of the theory \cite{Berges:2005hc}. In the present paper, we show that this result still holds for QED and, more generally, for any theory with linearly realized (gauge) symmetry. The crucial point is that the 2PI Ward identity (\ref{eq:gaugeinv}) allows for more counterterms than in usual perturbation theory.
For the purpose of illustration, we consider first the two-loop approximation of 2PI QED, namely the first diagram of Fig.~\ref{fig:loopQED}, as it already exhibits all the relevant features of the renormalization program.  We discuss the generalization to higher-loop orders in Sec.~\ref{sec:higher}. 

The propagators for renormalized and bare fields are related by $D_R=Z_2^{-1}\,D$ and $G_R=Z_3^{-1}\,G$ and the standard counterterms are defined as $\delta Z_{2}=Z_{2}-1$ and similarly for $\delta Z_3$, $Z_2\,m=Z_0m_R=m_R+\delta m$, $Z_2Z_3^{1/2}e=Z_1e_R=e_R+\delta e$. From here on, we only refer to renormalized quantities and we drop the subscript $R$ for simplicity. In terms of renormalized quantities, the 2PI functional $\Gammatpi[D,G]$ has a similar form as in Eq.~(\ref{eq:2PI}), with the replacement
\beq
\label{eq:replacement}
 \Gamma_{\rm int}[D,G]\longrightarrow \Gamma^R_{\rm int}[D,G]\equiv\Gamma_{\rm int}[D,G]
 +\delta\Gamma_{\rm int}[D,G]\,,
\eeq
where $\delta\Gamma_{\rm int}[D,G]$ denotes the contribution due to counterterms. As stated above, it should contain all closed 2PI graphs with local counterterms of mass dimension $d_M\le 4$ consistent with the symmetry property (\ref{eq:gaugeinv}) and Lorentz invariance. At two-loop order, it reads
\begin{eqnarray}\label{eq:ct}
 \delta\Gamma_{\rm int}[D,G]&=&\int_x\left\{
 -\tr\Big[(i\delta Z_2\slashchar{\partial}_x -\delta m)D(x,y)\Big]
 +\frac{\delta Z_3}{2}\,\Big(g^{\mu\nu}\partial_x^2
 -\partial_x^\mu\partial_x^\nu\Big)G_{\mu\nu}(x,y)\right.\nn
 &&\qquad+\frac{\delta\lambda}{2}\,\partial_x^\mu\partial_x^\nu G_{\mu\nu}(x,y)
 +\frac{\delta M^2}{2}\,{G^\mu}_\mu(x,x)\nonumber\\
 &&\qquad\left.+\,\frac{\delta g_1}{8}\,{G^\mu}_\mu(x,x){G^\nu}_\nu(x,x)+
 \frac{\delta g_2}{4}\,G^{\mu\nu}(x,x)G_{\mu\nu}(x,x)\right\}_{\!\!y=x}\,,
\end{eqnarray}
where ${\rm tr}$ denotes the trace over Dirac indices. The counterterms $\delta Z_2$, $\delta Z_3$ and $\delta m$ are the analog of the corresponding  ones in perturbation theory.\footnote{Notice that the counterterm $\delta e$ only appears at higher orders in the 2PI loop-expansion (see Sec.~\ref{sec:higher}).} The extra counterterms $\delta\lambda$, $\delta M^2$, $\delta g_1$ and $\delta g_2$ have no analog in perturbation theory but are allowed by the gauge symmetry (\ref{eq:gaugeinv}). We shall see below that their role is to absorb non-transverse divergences in the photon two- and four-point functions. In the following, we keep the discussion as general as possible so that it can be easily extended to higher-loop orders.

%%%%%%%%%% Renormalization
\section{Renormalization}
\label{sec:renormalization}

We now proceed to the elimination of UV divergences in Eqs. (\ref{eq:motion1})-(\ref{eq:motion2}) for the propagators $\bar D$ and $\bar G$. The power counting analysis of divergent integrals is identical to the case of a theory with a self-interacting scalar field coupled to a fermion field via a Yukawa-like interaction. A complete discussion of renormalization in such a theory can be found in~\cite{Reinosa:2005pj}. Here, we directly transpose the results of this analysis to the case of QED and discuss the issues related with gauge symmetry. 

%%%%%%%%%%%%%%%   Four-photon vertex
\subsection{Four-photon vertex}

Following Ref.~\cite{Reinosa:2005pj}, one can show that Eqs.~(\ref{eq:motion1})-(\ref{eq:motion2}) resum, in particular, an infinite series of potentially divergent sub-diagrams with four boson (photon) legs. This series defines a four-photon function $\bar V$, see Eq.~(\ref{eq:BS}) below. In the exact theory, the latter actually coincides with the 1PI four-photon vertex. It is thus transverse and its superficial degree of divergence, a priori equal to zero, is instead negative (see e.g. \cite{Weinberg}). This means that potential divergences with four photon legs are in fact absent. In contrast, at any finite order in the 2PI loop-expansion, transversality is only approximately fulfilled, which gives rise to actual divergent four-photon sub-diagrams. As we now show, these divergences can be absorbed in the counterterms $\delta g_1$ and $\delta g_2$ introduced above. 

%%%%%%%%%%   Definitions
\subsubsection{Definitions}\label{sec:definitions}

We define the 2PI four-point kernels $\bar\Lambda_{GG}$, $\bar\Lambda_{GD}$, $\bar\Lambda_{DG}$ and $\bar\Lambda_{DD}$ by\footnote{Note that we use a different sign convention as in Ref.~\cite{Reinosa:2005pj} for the definition of $\bar \Lambda_{DD}$.}
\begin{eqnarray}
\label{eq:kernels}
 \bar\Lambda_{GG}^{\mu\nu,\rho\sigma}(p,k) & \equiv & 
 \left.4\frac{\delta^2\Gamma^R_{\rm int}[D,G]}
 {\delta G_{\nu\mu}(p)\,\delta G_{\rho\sigma}(k)}\right|_{\bar D,\bar G}\,,
 \\
\label{eq:kernels2}
 \bar\Lambda_{GD}^{\mu\nu;\alpha\beta}(p,k) & \equiv & 
 \left.-2\frac{\delta^2\Gamma^R_{\rm int}[D,G]}
 {\delta G_{\nu\mu}(p)\,\delta D_{\alpha\beta}(k)}\right|_{\bar D,\bar G}\,,
 \\
\label{eq:kernels3}
 \bar\Lambda_{DG}^{\alpha\beta;\mu\nu}(p,k) & \equiv & 
 \left.-2\frac{\delta^2\Gamma^R_{\rm int}[D,G]}
 {\delta D_{\beta\alpha}(p) \,\delta G_{\mu\nu}(k)}\right|_{\bar D,\bar G}\,,
 \\
\label{eq:kernels4}
 \bar\Lambda_{DD}^{\alpha\beta,\delta\gamma}(p,k) & \equiv & 
 \left.\frac{\delta^2\Gamma^R_{\rm int}[D,G]}
 {\delta D_{\beta\alpha}(p)\,\delta D_{\delta\gamma}(k)}\right|_{\bar D,\bar G}\,.
\end{eqnarray}
Notice that the kernel $\bar\Lambda_{GG}$ receives an infinite constant contribution from the counterterms $\delta g_1$ and $\delta g_2$: 
\beq
\label{eq:GGkernel}
 \bar\Lambda_{GG}^{\mu\nu,\rho\sigma}(p,k) =  
\left.4\frac{\delta^2\Gamma_{\rm int}[D,G]}
 {\delta G_{\nu\mu}(p)\,\delta G_{\rho\sigma}(k)}\right|_{\bar D,\bar G}+\delta g_1
 \,g^{\mu\nu}g^{\rho\sigma}+\delta g_2\,\left(g^{\mu\rho}g^{\nu\sigma}+g^{\mu\sigma}g^{\nu\rho}\right)\,.
\eeq

The divergent sub-diagrams with four photon legs in $\bar D$ and $\bar G$ sum up to a function $\bar V^{\mu\nu,\rho\sigma}(p,k)$, which satisfies the following ladder equation~\cite{Reinosa:2005pj}:
\begin{eqnarray}\label{eq:BS}
 \bar V^{\mu\nu,\rho\sigma}(p,k) & = & 
 \bar K^{\mu\nu,\rho\sigma}(p,k)+\frac{1}{2}\int_q 
 \bar K^{\mu\nu,\bar\mu\bar\nu}(p,q)M^{GG}_{\bar\mu\bar\nu,\bar\rho\bar\sigma}(q)
 \bar V^{\bar\rho\bar\sigma,\rho\sigma}(q,k)\,,\nonumber\\
 & = &
 \bar K^{\mu\nu,\rho\sigma}(p,k)+\frac{1}{2}\int_q
 \bar V^{\mu\nu,\bar\mu\bar\nu}(p,q)M^{GG}_{\bar\mu\bar\nu,\bar\rho\bar\sigma}(q)
 \bar K^{\bar\rho\bar\sigma,\rho\sigma}(q,k)\,,
\end{eqnarray}
where $M^{GG}_{\mu\nu,\rho\sigma}(q)\equiv \bar G_{\mu\rho}(q)\bar G_{\sigma\nu}(q)$ and $\int_q\equiv i \int \frac{d^dq}{(2\pi)^d}$. The kernel $\bar K$ of this integral equation is defined as
\bea
\label{eq:Lambda_b}
 &&\bar K^{\mu\nu,\rho\sigma}(p,k) =
 \bar \Lambda_{GG}^{\mu\nu,\rho\sigma}(p,k)
 -\int_q \bar \Lambda_{GD}^{\mu\nu;\alpha\beta}(p,q)M^{DD}_{\alpha\beta,\gamma\delta}(q)
 \bar \Lambda_{DG}^{\gamma\delta;\rho\sigma}(q,k)\nn
 &&\qquad+
 \int_q\int_r\bar\Lambda_{GD}^{\mu\nu;\alpha\beta}(p,q)M^{DD}_{\alpha\beta,\bar\alpha\bar\beta}(q)
 \bar \Lambda^{\bar\alpha\bar\beta,\bar\gamma\bar\delta}(q,r)
 M^{DD}_{\bar\gamma\bar\delta,\gamma\delta}(r)\bar \Lambda_{DG}^{\gamma\delta;\rho\sigma}(r,k)\,,
\eea
where $M^{DD}_{\alpha\beta,\gamma\delta}(q)\equiv\bar D_{\alpha\gamma}(q)\bar D_{\delta\beta}(q)$, and where the function $\bar \Lambda(p,k)$ satisfies the ladder equation
\begin{eqnarray}
 \bar \Lambda^{\alpha\beta,\gamma\delta}(p,k) & = &
 \bar \Lambda_{DD}^{\alpha\beta,\gamma\delta}(p,k)
 -\int_q\bar\Lambda_{DD}^{\alpha\beta,\bar\alpha\bar\beta}(p,q)
 M^{DD}_{\bar\alpha\bar\beta,\bar\gamma\bar\delta}(q)
 \bar\Lambda^{\bar\gamma\bar\delta,\gamma\delta}(q,k)\,,\nonumber\\
 & = &
 \bar\Lambda_{DD}^{\alpha\beta,\gamma\delta}(p,k)
 -\int_q\bar\Lambda^{\alpha\beta,\bar\alpha\bar\beta}(p,q)
 M^{DD}_{\bar\alpha\bar\beta,\bar\gamma\bar\delta}(q)
 \bar\Lambda_{DD}^{\bar\gamma\bar\delta,\gamma\delta}(q,k)\,.
\label{eq:Lambda_f}
\end{eqnarray}
We show in App.~\ref{app:exact} that the set of Eqs.~(\ref{eq:BS})-(\ref{eq:Lambda_f}) actually define the 1PI four-photon vertex function. We finally note, for later use, that
\begin{equation}\label{eq:sym_id}
 \bar V^{\mu\nu,\rho\sigma}(p,k)=\bar V^{\nu\mu,\rho\sigma}(-p,k)
 =\bar V^{\mu\nu,\sigma\rho}(p,-k)=\bar V^{\sigma\rho,\nu\mu}(k,p)\,,
\end{equation}
and similarly for the functions $\bar\Lambda_{GG}$ and $\bar K$. 

%%%%%%%%%%   Renormalization of $\bar V^{\mu\nu,\rho\sigma}(p,k)$
\subsubsection{Renormalization of $\bar V^{\mu\nu,\rho\sigma}(p,k)$}

Here, we analyze the divergences encoded in Eqs.~(\ref{eq:BS})-(\ref{eq:Lambda_f}) and show how they can be absorbed in the counterterms $\delta g_{1,2}$ in (\ref{eq:ct}). We first note that these equations involve the propagators $\bar D$ and $\bar G$, which a priori bring their own divergences. However, as will be shown in Sec.~\ref{sec:propagators}, once the counterterms $\delta g_{1,2}$ have been used to renormalize the four-photon function $\bar V$, the propagators can be made finite by means of the remaining counterterms in Eq.~(\ref{eq:ct}). Thus, for the present discussion, it is sufficient to assume that $\bar D$ and $\bar G$ have been renormalized. Furthermore, we note that, at two-loop order, the kernels $\bar\Lambda_{GD}$, $\bar\Lambda_{DG}$ and $\bar\Lambda_{DD}$ are finite (since tree level). As we will see in Sec.~\ref{sec:higher}, this remains true at higher orders, provided one properly generalizes the shift $\delta\Gamma_{\rm int}[D,G]$ in \Eqn{eq:replacement}. Similarly, this allows one to remove all sub-divergences in $\bar\Lambda_{GG}$. 

To construct the four-photon vertex $\bar V$, one first resums the infinite series of ladder diagrams with rungs given by the kernel $\bar\Lambda_{DD}$ through Eq.~(\ref{eq:Lambda_f}). The resummed diagrams all have superficial degree of divergence $\delta=-2$ and a simple inspection shows that all possible sub-diagrams also have negative degree of divergence. It follows that the infinite series of ladder diagrams $\bar\Lambda$ is finite. The function $\bar\Lambda$ is then combined with the 2PI kernels $\bar\Lambda_{GG}$, $\bar\Lambda_{GD}$ and $\bar\Lambda_{DG}$ to construct the function $\bar K$, through Eq.~(\ref{eq:Lambda_b}). The corresponding diagrams all have superficial degree of divergence $\delta=0$. Here again, one can show by inspection that all possible sub-diagrams have negative degree of divergence. Therefore, Eq.~(\ref{eq:Lambda_b}) for $\bar K$ only contains overall logarithmic divergences, with tensor structure governed by the property (\ref{eq:sym_id}) and Lorentz symmetry:
\begin{equation}\label{eq:div4}
\Big[\bar K^{\mu\nu,\rho\sigma}\Big]_{\rm overall\,div}=\alpha_1\,g^{\mu\nu}g^{\rho\sigma} +\alpha_2\,\left(g^{\mu\rho}g^{\nu\sigma}+g^{\mu\sigma}g^{\nu\rho}\right)\,,
\end{equation}
with $\alpha_{1,2}$ two infinite constants. Finally, $\bar V$ is obtained from Eq.~(\ref{eq:BS}) as the infinite series of ladder diagrams with rungs given by $\bar K$. To identify divergences in $\bar V$, it is convenient to first consider the case where the counterterms $\delta g_{1,2}$ are set to zero, see Eq.~(\ref{eq:GGkernel}). This defines the functions $\bar K_{\delta g=0}$ and $\bar V_{\delta g=0}$.
The overall divergence of a ladder $L_{\rm r}$ with $r$ rungs in $\bar V_{\delta g=0}$ has the same Lorentz structure as above:
\begin{equation}\label{eq:overalls}
\Big[L_{\rm r}^{\mu\nu,\rho\sigma}\Big]_{\rm overall\, div}=\alpha^{(r)}_1\,g^{\mu\nu}g^{\rho\sigma} +\alpha^{(r)}_2\,\left(g^{\mu\rho}g^{\nu\sigma}+g^{\mu\sigma}g^{\nu\rho}\right)\,,
\end{equation}
with $\alpha^{(1)}_{1,2}=\alpha_{1,2}$. In addition, ladder diagrams also contain sub-divergences, again with the same Lorentz structure. 

\begin{figure}[t]
\begin{center}
\includegraphics[width=11cm]{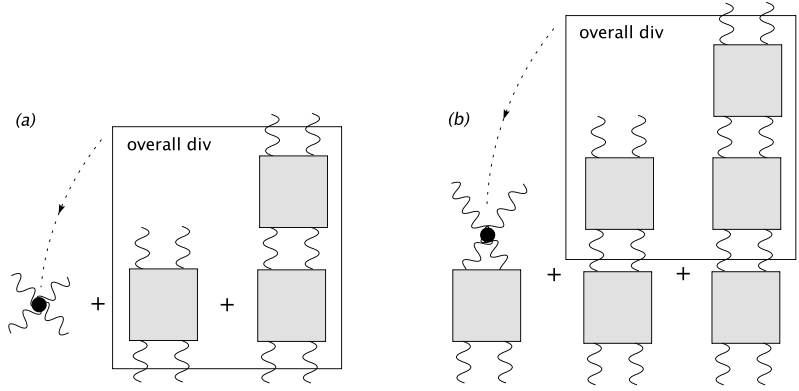}
\caption{\label{fig:BS} Self consistent renormalization of $\bar V$: overall divergences are absorbed in the tree level diagram containing $\delta g_{1,2}$ (a); ladder diagrams containing $\delta g_{1,2}$ serve to remove sub-divergences in the series of ladder diagrams (b). Grey boxes denote the function $\bar K$ with $\delta g_1$ and $\delta g_2$ set to~zero.}
\end{center}
\end{figure}
Taking now the contributions from the counterterms $\delta g_{1,2}$ in \Eqn{eq:GGkernel} into account, one sees that \Eqn{eq:BS} generates new ladder diagrams, which can be obtained from the previous ones by replacing an arbitrary number of rungs by the counterterm contribution $\delta g_1\,g^{\mu\nu}g^{\rho\sigma}+\delta g_2\,(g^{\mu\rho}g^{\nu\sigma}+g^{\mu\sigma}g^{\nu\rho})$.
The role of these diagrams is precisely to absorb overall divergences as well as sub-divergences, as we now illustrate. The tree level diagram involving $\delta g_{1,2}$ (originating from the ladder with a single rung $\bar K$), allows one to absorb all the overall divergences $\alpha^{(r)}_{1,2}$ by setting\footnote{Notice that this is an implicit equation for $\delta g_{1,2}$ since the overall divergences $\alpha^{(r)}_{1,2}$ actually depend on these counterterms, see below.}
\begin{equation}
\delta g_{1,2}=-\sum_{r=1}^\infty\alpha^{(r)}_{1,2}+{\rm finite}\,.
\end{equation}
This is illustrated in Fig.~\ref{fig:BS}.a. One can then show, following previous analysis in scalar theories~\cite{vanHees:2001ik,Berges:2005hc}, that the role of higher ladder diagrams containing $\delta g_{1,2}$ is to properly absorb all sub-divergences generated by the ladder resummation (including overlapping sub-divergences). This is illustrated in Fig. \ref{fig:BS}.b. Thus, the counterterms $\delta g_{1,2}$ account simultaneously for sub-divergences and overall divergences in Eq.~(\ref{eq:BS}). This is a consequence of the self-consistent character of the 2PI approximation scheme.

%%%%%%%%%%   Renormalization conditions
\subsubsection{Renormalization conditions} 
\label{sec:ren_conditions}

In order to fix the counterterms $\delta g_{1,2}$, one needs to impose appropriate renormalization conditions. We know that $\delta g_1=\delta g_2=0$ in the exact theory. Therefore, it is crucial to impose renormalization conditions which guarantee that $\delta g_{1,2}$ vanish in the limit $L\to\infty$. In order not to introduce new parameters or spurious constraints in the theory, it is natural to choose conditions which correspond to exact identities and are, therefore, automatically fulfilled in the exact theory.

A simple choice is the transversality condition (we choose a symmetric renormalization point for simplicity):
\begin{equation}\label{eq:renormalization_condition}
 P_{L,\mu\nu}(k_*)\bar V^{\mu\nu,\rho\sigma}(k_*,k_*)=0\,,
\end{equation}
where $P_L^{\mu\nu}(k_*)=n_*^\mu\,n_*^\nu$ is the longitudinal projector, with $n_*^\mu\equiv
k_*^\mu/\sqrt{k_*^2}$. As we now show, this condition guarantees that $\delta g_{1,2}={\cal O}(e^{2L})$. For this purpose, it is convenient to consider for a moment the bare function $\bar V_B$, that is the function $\bar V$ where all counterterms have been set to zero. In the exact theory, it coincides with the bare four-photon vertex (see App.~\ref{app:exact}) and is, therefore, transverse.\footnote{Here, it is crucial to employ an appropriate regularization scheme, such as e.g. dimensional regularization.} Moreover, at $L$-loop order in the 2PI loop-expansion, $\bar V_B$ agrees with the perturbative expansion of the exact bare four-photon vertex up to, and including, order $e^{2(L-1)}$. This is because the neglected higher-loop contributions to the 2PI effective action contribute to the kernels (\ref{eq:kernels})-(\ref{eq:kernels4}), and in turn to $\bar V_B$, starting at order $e^{2L}$. Thus the bare function $\bar V_B$ is transverse up to this order. This remains true for the sum of ladders $\bar V_{\delta g=0}$, defined in the previous subsection, which can be obtained from $\bar V_B$ after renormalizing the propagators $\bar D$ and $\bar G$, as well as the kernels $\bar\Lambda_{GD}$, $\bar\Lambda_{DG}$ and $\bar\Lambda_{DD}$, and removing sub-divergences in the kernel $\bar\Lambda_{GG}$. As we show later, the counterterms needed for these purposes bring non-transverse contributions to the function $\bar V_{\delta g=0}$, however only at order $e^{2L}$. It follows that $\bar V_{\delta g=0}$ is transverse up to this order. In particular, one has:
\begin{equation}\label{eq:renormalization_condition2}
 P_{L,\mu\nu}(k_*)\bar V^{\mu\nu,\rho\sigma}_{\delta g=0}(k_*,k_*)={\cal O}(e^{2L})\,.
\end{equation}
Subtracting \Eqn{eq:renormalization_condition2} from \Eqn{eq:renormalization_condition} and using \Eqn{eq:BS} as well as the fact that the function $\bar K_{\delta g=0}$ is ${\cal O}(e^4)$, one easily obtains that
\begin{equation}
 \delta g_1\,P_T^{\mu\nu}(k_*)
 +(\delta g_1+2\delta g_2)\,P_L^{\mu\nu}(k_*)+\mathcal{O}(e^4\,\delta g_{1,2})
 =\mathcal{O}(e^{2L})\,,
\end{equation}
where $P_T^{\mu\nu}(k_*)=g^{\mu\nu}-n_*^\mu n_*^\nu$. It follows that $\delta g_{1,2}=\mathcal{O}(e^{2L})$, as announced.\footnote{An explicit calculation of the divergent part of the counterterms $\delta g_{1,2}$ in the 2PI two-loop approximation is provided in App.~\ref{app:e4}.}

In practice, using \Eqn{eq:sym_id}, the general tensor structure of the four-photon vertex can be written as
\begin{eqnarray}
 \bar V^{\mu\nu,\rho\sigma}(k_*,k_*) =
 \bar{V}_1^*\,g^{\mu\nu}g^{\rho\sigma}
 +\bar{V}_2^*\left[g^{\mu\rho}g^{\nu\sigma}+g^{\mu\sigma}g^{\nu\rho}\right]
 +\bar{V}_3^*\left[g^{\mu\nu}n_*^\rho n_*^\sigma+ n_*^\mu n_*^\nu g^{\rho\sigma}\right]
 &&\nn
 +\bar{V}_4^*\left[g^{\mu\rho}n_*^\nu n_*^\sigma+ n_*^\mu n_*^\rho g^{\nu\sigma}\right]
 +\bar{V}_5^*\,g^{\mu\sigma}n_*^\nu n_*^\rho+ \bar{V}_6^*\,n_*^\mu n_*^\sigma g^{\nu\rho}
 +\bar{V}_7^*\,n_*^\mu n_*^\nu n_*^\rho n_*^\sigma,
 &&
\end{eqnarray}
where $\bar V_i^*\equiv \bar V_i(k_*^2)$. The renormalization condition (\ref{eq:renormalization_condition}) implies that
\begin{equation}
\bar{V}_1^*+\bar{V}_3^*=0 \quad \mbox{and} \quad \bar{V}_1^*+2\bar{V}_2^*+2\bar{V}_3^*+2\bar{V}_4^*+\bar{V}_5^*+\bar{V}_6^*+\bar{V}_7^*=0\,.
\end{equation}
These two equations fix the values of the independent counterterms $\delta g_1$ and $\delta g_2$ at each order in the 2PI loop-expansion. Finally, following the same steps as in scalar theories, one 
can write an explicitly finite equation, which makes no reference to the infinite counterterms
$\delta g_1$ and $\delta g_2$ anymore:
\begin{eqnarray}
\label{eq:Vfinite}
 && \bar V^{\mu\nu,\rho\sigma}(p,k) - \bar V^{\mu\nu,\rho\sigma}(k_*,k_*)
 =\bar K^{\mu\nu,\rho\sigma}(p,k)-\bar K^{\mu\nu,\rho\sigma}(k_*,k_*)\nonumber\\
 && \hspace{1.4cm}+\,\frac{1}{2}\int_q
 \bar V^{\mu\nu,\bar\mu\bar\nu}(p,q)\,
 M^{GG}_{\bar\mu\bar\nu,\bar\rho\bar\sigma}(q)\,
 \Big\{\bar  K^{\bar\rho\bar\sigma,\rho\sigma}(q,k)
 -\bar K^{\bar\rho\bar\sigma,\rho\sigma}(q,k_*)\Big\}\nonumber\\
 &&  \hspace{1.4cm}+\frac{1}{2}\int_q
 \Big\{\bar K^{\mu\nu,\bar\mu\bar\nu}(p,q)
 -\bar K^{\mu\nu,\bar\mu\bar\nu}(k_*,q)\Big\}\,
 M^{GG}_{\bar\mu\bar\nu,\bar\rho\bar\sigma}(q)\,
 \bar V^{\bar\rho\bar\sigma,\rho\sigma}(q,k_*)\,.
\end{eqnarray}
Indeed, the difference $\bar K^{\mu\nu,\bar\mu\bar\nu}(p,q)-\bar K^{\mu\nu,\bar\mu\bar\nu}(p',q')$
is finite, see \Eqn{eq:div4}, and it can be shown, using general topological properties of the the function
$\bar K$ and Weinberg theorem~\cite{Reinosa:2005pj}, that the differences between brackets under the 
integrals in Eq.~(\ref{eq:Vfinite}) decrease as $1/q$ at large $q$. Since $M^{GG}(q)\sim 1/q^4$ and 
$\bar V(p,q)\sim\ln q$ at large $q$ and fixed $p$, this guarantees the convergence of the
integrals.

%%%%%%%%%%%%%%%   Propagators
\subsection{Propagators}
\label{sec:propagators}
After $\delta g_{1,2}$ have be tunned to renormalize the four-photon function $\bar V$, only two-point divergences are left in the fermion and photon propagators. As we now show, these can be absorbed in the remaining counterterms in~\Eqn{eq:ct}. 

%%%%%%%%%%   General proparties of $\bar\Pi^{\mu\nu}(k)$
\subsubsection{General properties of $\bar\Pi^{\mu\nu}(k)$}
It is useful to recall some general properties of the photon polarization tensor (\ref{eq:motion2}). Lorentz invariance imposes the general structure
\beq
 \bar\Pi^{\mu\nu}(k)=P_T^{\mu\nu}(k)\,\bar\Pi_T(k^2)+P_L^{\mu\nu}(k)\,\bar\Pi_L(k^2)\,.
\eeq
The photon polarization tensor being a one-particle-irreducible structure, it is expected not to have any pole at $k^2=0$,\footnote{This assumes no spontaneous symmetry breaking, see e.g.~\cite{Weinberg}.} which implies that $\bar\Pi_T(0)=\bar\Pi_L(0)$ is a finite number. In the exact theory, the 1PI Ward identities ensure that the longitudinal part of the photon propagator is not modified by interactions: $\bar\Pi_L(k^2)=0$ for all $k^2$. It follows that $\bar\Pi_T(0)=0$, which guarantees that the photon mass is zero. However, $\bar\Pi^{\mu\nu}(k)$ is only approximately transverse at any finite order in the 2PI loop-expansion. At $L$-loop order, one has $\bar\Pi_L(k^2)\sim{\cal O}(e^{2L})$ and, therefore, $\bar\Pi_T(0)\sim{\cal O}(e^{2L})$. Hence, the pole of the photon propagator $\bar G^{\mu\nu}(k)$ is located at $k^2\sim{\cal O}(e^{2L})$. This is not a problem in principle since this stays within perturbative accuracy and can be systematically improved. The crucial issue is that non-transverse contributions to $\bar \Pi^{\mu\nu}(k)$ bring new possible type of divergences, as compared to the standard perturbative expansion. One has to check whether the latter can be removed with the counterterms~(\ref{eq:ct}) allowed by gauge invariance.

%%%%%%%%%%   Divergences
\subsubsection{Divergences and counterterms}
In terms of renormalized quantities, the fermion and photon self-energies read
\bea
\label{eq:sigmaren}
 \bar\Sigma_{\alpha\beta} (k)&=&
 -i\left.\frac{\delta \Gamma_{\rm int}[D,G]}{\delta D_{\beta\alpha}(k)}\right|_{\bar D,\bar G}
 +i\left(\delta Z_2\,\slashchar{k}-\delta m\right)_{\alpha\beta}\,.\\
\label{eq:piren}
 \bar\Pi^{\mu\nu}(k)&=&
 2i\left.\frac{\delta \Gamma_{\rm int}[D,G]}{\delta G_{\nu\mu}(k)}\right|_{\bar D,\bar G}
 -i\delta Z_3\,(g^{\mu\nu}k^2-k^\mu k^\nu)\nn
 &&+i\left[\delta M^2\, g^{\mu\nu}-\delta\lambda\, k^\mu k^\nu
 +\frac{\delta g_1}{2}\, g^{\mu\nu}{T^\rho}_\rho+\delta g_2\,T^{\mu\nu}\right]\,,
\eea
where $T^{\mu\nu}\equiv\int\frac{d^dq}{(2\pi)^d}\bar G^{\mu\nu}(q)$. After $\delta g_1$ and $\delta g_2$ have been fixed, 
there remain overall divergences in the self-energies (\ref{eq:sigmaren}) and (\ref{eq:piren}). For the
fermion self-energy, their Lorentz structure is $a+b \slashchar{k}$, with $a$ and $b$ two infinite 
constants which can be absorbed in the counterterms $\delta m$ and $\delta Z_2$. 
Similarly, the general structure of the remaining divergences in the photon self-energy reads $\left(\alpha+\beta k^2\right) g^{\mu\nu}+\gamma k^\mu k^\nu$, with $\alpha$, $\beta$ and $\gamma$ infinite constants which can be absorbed by the counterterm contribution
\beq
 (\delta M^2-\delta Z_3 k^2) g^{\mu\nu}+(\delta Z_3-\delta\lambda)k^\mu k^\nu\,.
\eeq
As in the case of the four-photon vertex, the counterterms needed to absorb the overall divergences in Eqs.~(\ref{eq:sigmaren}) and (\ref{eq:piren}) are just what is needed to actually get rid of two-point sub-divergences. Again, this is due to the self-consistent character of the 2PI approximation scheme.

We see here the importance of allowing for all counterterms permitted by the symmetry (\ref{eq:gaugeinv}) of the 2PI functional. In the exact theory, the photon polarization tensor is transverse. It follows that $\alpha=\beta+\gamma=0$. Therefore, all divergences can be absorbed in the single counterterm $\delta Z_3$ and there is no need for the counterterms $\delta M^2$ and $\delta\lambda$. However, this is not true at any finite order in the 2PI loop-expansion and it is crucial to take these counterterms into account. 

%%%%%%%%%%   Renormalization conditions
\subsubsection{Renormalization conditions}
The counterterms $\delta m$ and $\delta Z_2$ can be fixed by imposing the following usual renormalization conditions
\begin{equation}
 \bar{\Sigma}(k_*)=0 \qquad \mbox{and} \qquad 
 \left.\frac{d\bar{\Sigma}(k)}{d\slashchar{k}}\right|_{k=k_*}=0\,,
\end{equation}
where we used the same renormalization point as above for the four-photon vertex. 
Similarly, for the photon self-energy, one can impose the following condition on the transverse part of the photon polarization tensor:
\begin{equation}
 \left.\frac{d\bar{\Pi}_{\rm T}(k^2)}{dk^2}\right|_{k=k_*}=0\,.
\end{equation}
In perturbation theory, this is enough to fix the photon wave-function renormalization counterterm. In the 2PI framework, however, one needs to fix the three counterterms $\delta Z_3$, $\delta\lambda$ and $\delta M^2$ and two other independent conditions are needed. Again, we choose renormalization conditions which are automatically fulfilled in the exact theory in order to ensure that these counterterms converge to their correct values, in particular $\delta\lambda,\delta M^2\to~0$, in the limit $L\to\infty$. A possible choice is to impose transversality of the photon polarization tensor at the scale $k_*$:
\begin{equation}
\label{eq:transcond}
 \bar{\Pi}_{\rm L}(k^2_*)=0\,.
\end{equation}
Since in the exact theory the transversality condition has to hold for all momenta, one can further impose that
\begin{equation}
 \left.\frac{d\bar\Pi_{\rm L}(k^2)}{d k^2}\right|_{k=k_*}=0\,.
\end{equation}
Using similar arguments as before, one can show that the above renormalization conditions guarantee that, at $L$-loop order, $\delta M^2$ and $\delta\lambda$ are of order $e^{2L}$. This in turn shows that non-transversalities in $\bar V$, arising from the renormalization of the propagators appear only at order $e^{2L}$, as announced in Sec.~\ref{sec:ren_conditions}.

As a final comment, we mention that, although the above conditions at the renormalization scale $k=k_*$ are sufficient for renormalization, one might employ a physical renormalization point. For instance, it is interesting to note that imposing the transversality condition (\ref{eq:transcond}) at $k_*=0$ implies that $\bar\Pi_T(0)=\bar\Pi_L(0)=0$ and, therefore, that the pole of the photon propagator $\bar G_{\mu\nu}(k)$ is exactly located at $k^2=0$ order by order in the 2PI loop-expansion. 

%%%%%%%%%%%%%%%%%%%%   Higher loops
\section{Higher loops}
\label{sec:higher}

The previous analysis is pretty general and only requires the kernels $\bar\Lambda_{GD}$, $\bar\Lambda_{DG}$, $\bar\Lambda_{DD}$ to be finite as well as $\bar\Lambda_{GG}$ to only contain overall divergences. Since $\bar\Lambda_{GD}$, $\bar\Lambda_{DG}$, $\bar\Lambda_{DD}$ have a negative superficial degree of divergence, one can formulate this condition in a more symmetric manner by saying that the 2PI-kernels \eqn{eq:kernels}-\eqn{eq:kernels4} have to be void of sub-divergences. Higher loop contributions to $\Gamma_{\rm int}[D,G]$ come with corresponding counterterms in $\delta\Gamma_{\rm int}[D,G]$. Again, at any given order, one should include all possible 2PI diagrams containing local counterterms of mass dimension $d_M\le 4$ allowed by the gauge symmetry (\ref{eq:gaugeinv}). Below, we classify possible counterterms which can appear in higher-loop diagrams and show that they precisely provide what is needed to make the kernels \eqn{eq:kernels}-\eqn{eq:kernels4} fulfill the above mentioned prerequisites (see also \cite{vanHees:2001ik,Berges:2005hc,Reinosa:2005pj}).

Since only 2PI diagrams contribute to the effective action, it is clear that there can be no two-point counterterms --~that is counterterms corresponding to a two-point vertex~-- in diagrams with more than one loop~\cite{Berges:2005hc}, i.e. beyond those already present in the first two lines of Eq.~(\ref{eq:ct}). Therefore, these counterterms never appear explicitly in the expressions of the kernels \eqn{eq:kernels}-\eqn{eq:kernels4}. Notice that for the same topological reason, there can be no two-point sub-structure -- and hence no two-point sub-divergence -- in these kernels, beyond those implicitly hidden in the propagators. In fact the only role of these counterterms is to absorb two-point divergences in the propagators $\bar D$ and $\bar G$, which the kernels \eqn{eq:kernels}-\eqn{eq:kernels4} are made of.

\begin{figure}[t]
\begin{center}
\includegraphics[width=7cm]{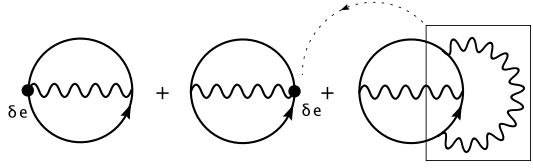}
\caption{\label{fig:charge} The complete three-loop contributions to $\Gamma_{\rm int}^R[D,G]$.
The box shows a three-point structure which eventually leads to a divergent correction to the QED vertex in the kernels (4.1)-(4.4), to be absorbed in the charge counterterm $\delta e$.}
\end{center}
\end{figure}
In contrast, explicit three-point sub-diagrams with one photon leg and two fermion legs appear in the kernels \eqn{eq:kernels}-\eqn{eq:kernels4} beyond the 2PI two-loop approximation. They have superficial degree of divergence zero and correspond to charge renormalization. As an illustration, consider the three-loop contribution to the 2PI effective action represented in Fig.~\ref{fig:phi}, which reads:
\beq
\label{eq:threeloop}
 \Gamma_{\rm int}^{\rm (3-loop)}[D,G]=
 -\frac{1}{4}\int_{x_1\ldots x_4}{\cal F}^{\mu\nu\rho\sigma}(x_1,\ldots,x_4)\,
 G_{\mu\rho}(x_1,x_3)\,G_{\nu\sigma}(x_2,x_4) \,,
\eeq
where we defined the fermion loop as
\beq
\label{eq:fermionloop}
 {\cal F}^{\mu\nu\rho\sigma}(x_1,\ldots,x_4)\equiv 
 -ie^4\tr[\gamma^\mu D(x_1,x_2)\gamma^\nu D(x_2,x_3)
 \gamma^\rho D(x_3,x_4)\gamma^\sigma D(x_4,x_1)].
\eeq
Clearly, this does not bring any sub-divergence to the kernel $\bar\Lambda_{GG}$, but merely modifies its overall divergence (see App.~\ref{app:e4}). In contrast, it gives rise to sub-divergences in the kernels $\bar\Lambda_{GD}$, $\bar\Lambda_{DG}$ and $\bar\Lambda_{DD}$, which can be eliminated by means of charge renormalization. At the level of the 2PI functional, this corresponds to adding the following contribution to the shift $\delta\Gamma_{\rm int}[G,D]$:
\beq
\label{eq:threeloopshift}
 \delta\Gamma_{\rm int}^{\rm (3-loop)}[G,D]=
 -ie\,\delta e\int_{x_1x_2}\tr\left[\gamma^\mu D(x_1,x_2)\gamma^\nu D(x_2,x_1)\right]
 G_{\mu\nu}(x_1,x_2) \,,
\eeq
where $\delta e$ is the charge counterterm.
The sum of the three-loop contributions (\ref{eq:threeloop}) and (\ref{eq:threeloopshift}) to $\Gamma_{\rm int}^R[D,G]$ is represented diagrammatically in Fig. \ref{fig:charge}. Similarly, any higher-order diagram in $\Gamma_{\rm int}[D,G]$ containing sub-structures corresponding to QED vertex corrections must be accompanied with corresponding ones in $\delta\Gamma_{\rm int}[D,G]$ where these sub-structures are replaced by the local counterterm $-i\gamma^\mu\delta e$. 

\begin{figure}[t]
\begin{center}
\includegraphics[width=4cm]{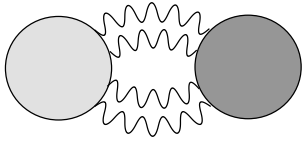}
\caption{\label{fig:4type2} General structure of diagrams in the 2PI functional leading to four-photon sub-divergences in the kernels (4.1)-(4.4). The blobs represent arbitrary four-photon sub-diagrams.}
\end{center}
\end{figure}
\begin{figure}[t]
\begin{center}
\includegraphics[width=12.8cm]{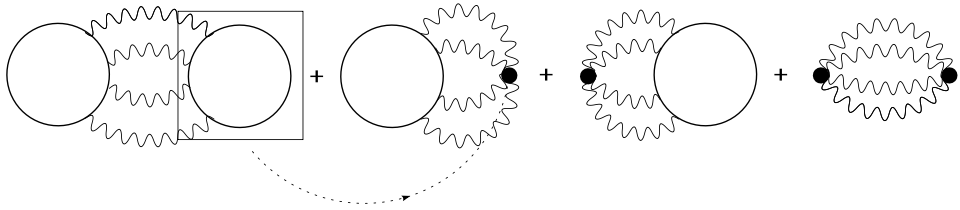}
\caption{\label{fig:four} Five-loop contributions to $\Gamma_{\rm int}[D,G]$ leading to four-photon sub-divergences in the 2PI-kernels (4.1)-(4.4) with the corresponding contributions to $\delta\Gamma_{\rm int}[D,G]$ involving four-photon counterterms. The loop with a solid line represents the one-loop four-photon diagram symmetrized with respect to its external legs.}
\end{center}
\end{figure}
\begin{figure}[t]
\begin{center}
\includegraphics[width=2.5cm]{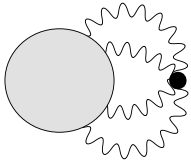}
\caption{\label{fig:structure} A general structure involving the four-photon counterterm $\delta g$ in $\Gamma_{\rm int}[D,G]$.}
\end{center}
\end{figure}

Finally, higher-loop contributions to the 2PI effective action also give rise to divergent four-photon sub-structures in the kernels \eqn{eq:kernels}-\eqn{eq:kernels4}. Such contributions have the general form depicted in Fig.~\ref{fig:4type2}. As an example, consider the case where the blobs in Fig.~\ref{fig:4type2} are both replaced by one fermion loop, as depicted on Fig.~\ref{fig:four}. The corresponding contribution to $\Gamma_{\rm int}[D,G]$ reads
\bea
 &&\frac{i}{48}\int_{x_1\ldots x_4}\int_{y_1,\ldots,y_4}{\cal F}_S^{\mu\nu\rho\sigma}(x_1,\ldots,x_4)\,
 G_{\mu\bar\mu}(x_1,y_1)\,G_{\nu\bar\nu}(x_2,y_2)\nn
\label{eq:5loop}
 &&\qquad\qquad\qquad\qquad\times G_{\rho\bar\rho}(x_3,y_3)\,G_{\sigma\bar\sigma}(x_4,y_4)\,
 {\cal F}_S^{\bar\mu\bar\nu\bar\rho\bar\sigma}(y_1,\ldots,y_4)\,,
\eea
where ${\cal F}_S$ is the symmetrized fermion loop: ${\cal F}_S^{\mu\nu\rho\sigma}(x_1,\ldots,x_4)={\cal F}^{\mu\nu\rho\sigma}(x_1,x_2,x_3,x_4)+{\cal F}^{\mu\nu\sigma\rho}(x_1,x_2,x_4,x_3)+\cdots$, see \Eqn{eq:fermionloop}. Clearly, the contribution \eqn{eq:5loop} leads to divergent one-loop sub-structures in the kernels $\bar\Lambda_{GG}$, $\bar\Lambda_{GD}$, $\bar\Lambda_{DG}$ and $\bar\Lambda_{DD}$. These have zero superficial degree of divergence and can be eliminated by means of a local four-photon counterterm $\delta g$. In the present case, this corresponds to adding the three counterterm diagrams represented on Fig.~\ref{fig:four}: the total contribution takes the same form as above, with the replacement\footnote{It is easy to see that four-photon sub-structure arising from the diagrammatic expansion of the 2PI functional beyond two-loop are symmetric in their Lorentz indices, hence the structure of the counterterm contribution.}
\beq
 {\cal F}_S^{\mu\nu\rho\sigma}(x_1,\ldots,x_4)\to
 {\cal F}_S^{\mu\nu\rho\sigma}(x_1,\ldots,x_4)+
 \delta g\,[g^{\mu\nu}g^{\rho\sigma}+g^{\mu\rho}g^{\nu\sigma}+g^{\mu\sigma}g^{\nu\rho}]\,.
\eeq
More generally, higher-loop diagrams in $\Gamma_{\rm int}[D,G]$ containing four-photon sub-structures must be accompanied with corresponding ones in $\delta\Gamma_{\rm int}[D,G]$ where each such sub-structure is replaced by a local counterterm $-i(\delta g/4!) [g^{\mu\nu} g^{\rho\sigma} + g^{\mu\rho} g^{\nu\sigma} + g^{\mu\sigma}g^{\nu\rho}]$, as illustrated in Fig.~\ref{fig:structure}.
Note that, at a given finite order in the 2PI loop-expansion, the actual values of the counterterms $\delta e$ and $\delta g$ differ for each graph where they appear, depending on the order of the graph itself. Just as charge counterterms, such four-photon counterterms are allowed by the gauge symmetry \eqn{eq:gaugeinv} and are necessary in order to obtain finite results from a 2PI calculation.

Finally, we need to check that, as announced earlier, diagrams containing four-photon counterterms in the kernels \eqn{eq:kernels}-\eqn{eq:kernels4}, which give rise to non-transverse terms in the function $\bar V$, only contribute at order $e^{2L}$ in a 2PI $L$-loop order calculation. To illustrate this, consider the sum of diagrams having the general structure represented in Fig.~\ref{fig:4type2}, where the left blob stands for a given $k$-loop sub-diagram. By construction, the right blob is nothing but the skeleton expansion of the exact 1PI four-photon vertex function at $\ell=L-k-3$ loops ($\ell \ge1$). Therefore, the corresponding four-photon sub-structure appearing in the kernels \eqn{eq:kernels}-\eqn{eq:kernels4} agrees with the exact 1PI four-photon vertex function up to, and including, order $e^{2(\ell+1)}$.\footnote{Remember that the propagators $\bar D$ and $\bar G$ agree with their respective perturbative expansions up to order $e^{2L}$.} It follows that the overall divergence of this sub-structure can be absorbed in a couterterm $\delta g = {\cal O}(e^{2(\ell+2)})$, which, therefore, vanishes in the exact theory, as it should. Moreover, since the four-photon counterterm considered here is involved in a diagram of order $e^{2(k+1)}$, see Fig.~\ref{fig:structure}, its contribution to the 2PI-kernels \eqn{eq:kernels}-\eqn{eq:kernels4} and, thereby, to the function $\bar V$ is indeed ${\cal O}(e^{2L})$.

\section{Conclusion}

We have shown how to renormalize the Schwinger-Dyson equations for the propagators derived from the 2PI effective action in QED in the linear covariant gauge. The elimination of UV divergences amounts to shifting the 2PI effective action by a counterterm part, given by the series of 2PI diagrams which contain all possible local counterterms of mass dimension less or equal to four allowed by the gauge symmetry of the 2PI functional. Besides the standard QED counterterms, this includes new countertems which have no analog in perturbation theory and whose role is to absorb divergences due to non-transverse contributions to the two- and four-photon vertex functions. By imposing suitable renormalization conditions, these counterterms are guaranteed to be systematically of higher order than the truncation order in a 2PI loop-expansion and to vanish in the exact theory. This guarantees the convergence of the approximation scheme towards the correct exact theory, described by the usual mass, charge and gauge fixing parameters $m$, $e$ and $\lambda$.

The present work opens the possibility of performing explicit (finite) calculations in 2PI QED. The renormalization of the 2PI-resummed effective action, which encodes all the vertices of the theory, along the lines of Ref.~\cite{Berges:2005hc}, will be presented elsewhere \cite{JU2}. It would be interesting to examine the relation between the present approach, which involves unusual counterterms, and functional renormalization group techniques involving modified Ward identities~\cite{Pawlowski:2005xe}.

Finally, it is to be mentioned that the analysis of divergences presented here applies to general gauge theories with arbitrary gauge fixing term since it only relies on power counting. We point out that the symmetry constraints, which, in particular, fix the structure of the shift $\delta \Gamma_{\rm int}$ needed to eliminate these divergences at each finite order in the 2PI loop-expansion, are particularly simple to solve for a theory with a linearly realized gauge symmetry. This is for instance the case of QCD quantized in the background field gauge~\cite{Weinberg}.

\section*{Acknowledgments}

We thank J. Berges, J.-P. Blaizot, Sz. Bors\'anyi and E. Iancu for fruitful collaboration on related topics. U. R. acknowledges support from the German Research Foundation (DFG), project no. WE 1056/4-4

%%%%%%%%%%%%%%%%%%%%   Appendix
\appendix

%%%%%%%%%%%%%%%%%%%%   Exact theory
\section{The four-photon vertex}
\label{app:exact}

Here, we show how the 1PI four-photon vertex can be obtained from the 2PI effective action in the exact theory. We follow closely the analog construction for scalar field theories given in \cite{Berges:2005hc}. Consider the following generating functional:
\beq 
 \e^{iW[H,R]}=\int{\mathcal D}A{\mathcal D}\psi{\mathcal D}\bar\psi\,
 \e^{i({\mathcal S}[A,\psi,\bar\psi]+
 \frac{1}{2}H_{ab}A_bA_a-R_{ab}\psi_b\bar\psi_a)}\,,
\eeq
where ${\mathcal S}[A,\psi,\bar\psi]$ is the classical action \eqn{eq:classact} and where $H$ and $R$ are bilinear sources. Throughout this section, the subscripts $a,b$ denote space-time variables as well as Lorentz or Dirac indices for each type of fields. The connected two-point functions $G$ and $D$ for the photon and fermion fields in the presence of sources can be obtained as
\beq
 \frac{\delta W}{\delta H_{21}}=\frac{1}{2}G_{12}\qquad {\rm and} \qquad
 \frac{\delta W}{\delta R_{21}}=-D_{12}\,.
\eeq
The 2PI effective action (\ref{eq:2PI}) can be defined as the double Legendre transform of the functional $W[H,R]$:
\beq
 \Gammatpi[D,G]=W[H,R]-H_{ab}\frac{\delta W}{\delta H_{ab}}-R_{ab}\frac{\delta W}{\delta R_{ab}}\,.
\eeq
One has, in particular,
\beq
 \frac{\delta \Gammatpi}{\delta G_{21}}=-\frac{1}{2}H_{12}\qquad {\rm and}\qquad
 \frac{\delta \Gammatpi}{\delta D_{21}}=R_{12}\,.
\eeq

Expressing the fact that the Jacobian of the double Legendre transformation is unity, one obtains the following relations:
\bea
\label{app:jacobian1}
 2\,\frac{\delta^2\Gammatpi}{\delta G_{21}\delta G_{ab}}\,
 \frac{\delta^2W}{\delta H_{ba}\delta H_{34}}
 -\frac{\delta^2\Gammatpi}{\delta G_{21}\delta D_{ab}}\,
 \frac{\delta^2W}{\delta R_{ba}\delta H_{34}}
 &=&-\frac{1}{2}\,1\!\!1_{12,34}\,,\\
\label{app:jacobian2}
 2\,\frac{\delta^2\Gammatpi}{\delta D_{21}\delta G_{ab}}\,
 \frac{\delta^2W}{\delta H_{ba}\delta H_{34}}
 -\frac{\delta^2\Gammatpi}{\delta D_{21}\delta D_{ab}}\,
 \frac{\delta^2W}{\delta R_{ba}\delta H_{34}}
 &=&0\,,
\eea
where summation over repeated indices is implied and where we defined $1\!\!1_{12,34}\equiv(\delta_{13}\delta_{24}+\delta_{14}\delta_{23})/2$. The second derivatives of the generating functional $W[H,R]$ are related to the four-point functions of the theory. In particular, one has
\bea
\label{app:con4}
 \frac{\delta^2W}{\delta H_{21}\delta H_{34}}&=&
 \frac{i}{4}\left(G_{13}G_{42}+G_{14}G_{32}+C^{(0,4)}_{1234}\right)\,,\\
 \frac{\delta^2W}{\delta R_{21}\delta H_{34}}&=&-\frac{i}{2}C^{(2,2)}_{1234}\,,
\eea
where the $C^{(n,m)}$'s are the connected $(n+m)$-point functions with $n$ fermion and $m$ photon legs. They are related to the corresponding amputated four-point functions $\Gamma^{(n,m)}$ as
\bea
\label{app:amp}
 C^{(0,4)}_{1234}&\equiv& i G_{1a}G_{b2}G_{3c}G_{d4}\,\Gamma^{(0,4)}_{abcd}\,,\\
 C^{(2,2)}_{1234}&\equiv& i D_{1a}D_{b2}G_{3c}G_{d4}\,\Gamma^{(2,2)}_{abcd}\,.
\eea
From the parametrization \eqn{eq:2PI} of the 2PI effective action, one has
\bea
\label{app:secder2pi1}
 \frac{\delta^2\Gammatpi}{\delta G_{21}\delta G_{34}}&=&
 \frac{i}{4}\Big(G_{13}^{-1}G_{42}^{-1}+G_{14}^{-1}G_{32}^{-1}\Big)
 +\frac{\delta^2\Gamma_{\rm int}}{\delta G_{21}\delta G_{34}}\,,\\
\label{app:secder2pi2}
 \frac{\delta^2\Gammatpi}{\delta D_{21}\delta D_{34}}&=&
 -iD_{13}^{-1}D_{42}^{-1}+\frac{\delta^2\Gamma_{\rm int}}{\delta D_{21}\delta D_{34}}\,,\\ 
\label{app:secder2pi3}
 \frac{\delta^2\Gammatpi}{\delta G_{21}\delta D_{34}}&=&
 \frac{\delta^2\Gamma_{\rm int}}{\delta G_{21}\delta D_{34}}\,.
\eea
Inserting Eqs.~\eqn{app:amp}-\eqn{app:secder2pi3} in Eqs.~\eqn{app:jacobian1}-\eqn{app:jacobian2} and using the definitions \eqn{eq:kernels}-\eqn{eq:kernels4}, one obtains the following set of coupled ladder equation for the amputated four-point functions, in the absence of sources (hence the bars):\footnote{Similar equations involving the amputated four-fermion function $\Gamma^{(4,0)}$ can be written, but they do not play any role for the purpose of renormalization.}
\bea
\label{app:gamma04}
 \bar\Gamma^{(0,4)}&=&\bar \Lambda_{GG}+{i\over2}\,\bar \Lambda_{GG}\,M^{GG}\,\bar\Gamma^{(0,4)}
 -i\bar \Lambda_{GD}\,M^{DD}\,\bar\Gamma^{(2,2)}\,,\\
\label{app:gamma22}
 \bar\Gamma^{(2,2)}&=&\bar \Lambda_{DG}+{i\over2}\,\bar \Lambda_{DG}\,M^{GG}\,\bar\Gamma^{(0,4)}
 -i\bar \Lambda_{DD}\,M^{DD}\,\bar\Gamma^{(2,2)}\,. 
\eea
where $M^{GG}_{ab,cd}\equiv\bar G_{ac}\bar G_{db}$, and similarly for $M^{DD}$. Here we left space-time/Lorentz/Dirac indices implicit for simplicity.

This set of equations is easily shown to lead to the defining equations \eqn{eq:BS}-\eqn{eq:Lambda_f} for the vertex $\bar V$, by explicitly eliminating $\bar\Gamma^{(2,2)}$. Indeed, Eq.~\eqn{app:gamma22} can be formally solved by means of the function $\bar\Lambda$, defined as the solution of the following ladder equation
\beq
\label{app:lambda}
 \bar\Lambda=\bar\Lambda_{DD}-i\bar\Lambda_{DD}\,M^{DD}\,\bar\Lambda\,,
\eeq
which resums the infinite series of ladder diagrams with rungs $\bar\Lambda_{DD}$. One has
\beq
\label{app:gam222}
 \bar\Gamma^{(2,2)}=A-i\bar\Lambda\,M^{DD}\,A\,,
\eeq
where we defined
\beq
 A \equiv \bar\Lambda_{DG}+{i\over2}\,\bar \Lambda_{DG}\,M^{GG}\,\bar\Gamma^{(0,4)}.
\eeq
Inserting Eq.~\eqn{app:gam222} into Eq.~\eqn{app:gamma04}, and defining the function $\bar K$ as
\beq
 \bar K \equiv \bar\Lambda_{GG}-i\bar\Lambda_{GD}\,M^{DD}\,\bar\Lambda_{DG}
 +i^2\bar\Lambda_{GD}\,M^{DD}\,\bar\Lambda\,M^{DD}\,\bar\Lambda_{DG}\,,
\eeq
one finally obtains the following equation for the amputated four-photon function:
\beq
\label{app:gam042}
 \bar\Gamma^{(0,4)}=\bar K+\frac{i}{2}\,\bar K\,M^{GG}\,\bar\Gamma^{(0,4)}\,.
\eeq

We note that all the equations in this section can be used interchangeably for either bare or renormalized quantities. We see that, in the exact theory, Eqs.~\eqn{app:lambda}-\eqn{app:gam042} for the bare four-photon vertex coincide with Eqs.~(\ref{eq:BS})-(\ref{eq:Lambda_f}) where all counterterms are set to zero, which defines the bare function $\bar V_B$ used in Sec.~\ref{sec:ren_conditions}.

%%%%%%%%%%%%%%%%%%%%   $e^4$ contributions to $\bar{V}^{\mu\nu,\rho\sigma}$
\section{Cancellation of four-photon divergences}
\label{app:e4}

In this section, we illustrate on an explicit example how four-photon divergences --~and hence the divergent part of the counterterms $\delta g_1$ and $\delta g_2$~-- are systematically pushed to ${\cal O}(e^{2L})$ in the 2PI loop-expansion. 
\begin{figure}[t]
\begin{center}
\includegraphics[width=13.5cm]{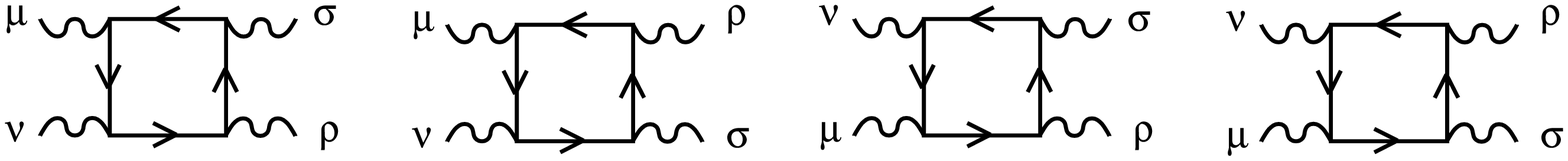}
\end{center}
\vspace*{-0.5cm}
\caption{$e^4$ contributions to $\bar{V}^{\mu\nu,\rho\sigma}(p,k)$ from the 2PI two-loop approximation. The momenta associated with each line are as follows: ($\mu$) incoming $p$ ; ($\nu$) outgoing $p$ ; ($\rho$) outgoing $k$ ; ($\sigma$) incoming $k$.}
\label{fig:e41}
\end{figure}

Consider the two-loop ($L=2$) approximation of the 2PI effective action, corresponding to the first diagram of Fig.~\ref{fig:phi}. Deriving the kernels \eqn{eq:kernels}-\eqn{eq:kernels4} and iterating them through the set of equations \eqn{eq:BS}-\eqn{eq:Lambda_f} to obtain the vertex $\bar{V}$, one easily obtains that, as expected, the latter starts at order $e^4$ with the box diagrams represented on Fig.~\ref{fig:e41}, where the lines represent the fermion propagator $\bar D$. To obtain the perturbative $e^4$-contribution, one replaces the propagator lines by free propagators. Moreover, for the purpose of calculating its UV divergence, it is enough to consider vanishing external momenta and to remove masses from the numerator. This is because the superficial degree of divergence of the diagram is $\delta=0$ and there are no sub-divergences. Consider the first of these diagrams. The potentially divergent integral reads, in Euclidean momentum space,
\begin{equation}
 I^{\mu\nu\rho\sigma}=-e^4\int \frac{d^dp}{(2\pi)^d}
 \frac{\mbox{tr}\,\big[\gamma^\mu\slashchar{p}\gamma^\nu\slashchar{p}\gamma^\rho
 \slashchar{p}\gamma^\sigma\slashchar{p}\big]}{(p^2+m^2)^4}\,,
\end{equation}
where we have kept the fermion mass in the denominator for IR safety. After some calculations, one obtains
\begin{equation}
 I^{\mu\nu\rho\sigma}=-e^4\,\Big[A\,\left(g^{\mu\nu}g^{\rho\sigma}
 +g^{\mu\sigma}g^{\nu\rho}\right)+B\,g^{\mu\rho}g^{\nu\sigma}\Big]\,I_{\rm div}\,,
\end{equation}
with
\begin{equation}
 A=4\,\frac{d-2}{d+2} \qquad \mbox{and} \qquad B=-(d+4)\,\frac{4}{d}\,\frac{d-2}{d+2}\,,
\end{equation}
and ($d=4-2\epsilon$)
\begin{equation}
\label{app:Idiv}
 I_{\rm div}=\int \frac{d^dp}{(2\pi)^d}\frac{(p^2)^2}{(p^2+m^2)^4}=\frac{1}{16\pi^2}\frac{1}{\epsilon}+\mbox{finite}\,.
\end{equation}
The complete two-loop contribution to order $e^4$ is then given by
\begin{equation}
I^{\mu\nu\rho\sigma}+I^{\mu\nu\sigma\rho}+I^{\nu\mu\rho\sigma}+I^{\nu\mu\sigma\rho}
=-2e^4\Big[2A\,g^{\mu\nu}g^{\rho\sigma}
+(A+B)\left(g^{\mu\rho}g^{\nu\sigma}+g^{\mu\sigma}g^{\nu\rho}\right)\Big]\,I_{\rm div}\,.
\end{equation}
For $d\rightarrow 4$, the pre-factor multiplying the divergent integral is $\propto 2\,g^{\mu\nu}g^{\rho\sigma}-g^{\mu\sigma}g^{\nu\rho}-g^{\mu\rho}g^{\nu\sigma}\neq 0$. Thus there are non-vanishing four-photon divergences at order $e^4$ in the 2PI two-loop approximation. 
\begin{figure}[t]
\begin{center}
\includegraphics[width=6.4cm]{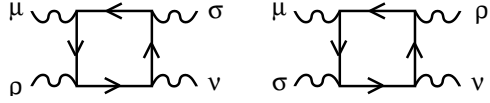}
\end{center}
\vspace*{-0.5cm}
\caption{$e^4$ contributions to $\bar{V}^{\mu\nu,\rho\sigma}(p,k)$ from the 2PI three-loop approximation. Momenta are organized as in Fig.~7.}
\label{fig:e42}
\end{figure}

How are these divergences pushed to order $e^6$ when one goes to three-loop? The three-loop contribution to the 2PI functional $\Gamma_{\rm int}[G,D]$ is shown on Fig.~\ref{fig:phi}. It gives a new $e^4$ contribution to the four-photon vertex $\bar V$, represented on Fig.~\ref{fig:e42}, which is precisely what is needed to make the total $e^4$ contribution symmetric and transverse, and hence finite. Indeed, the total $e^4$ contribution is now given by
\begin{equation}
 I^{\mu\nu\rho\sigma}+I^{\mu\nu\sigma\rho}+I^{\nu\mu\rho\sigma}
 +I^{\nu\mu\sigma\rho}+I^{\mu\rho\nu\sigma}+I^{\mu\sigma\nu\rho}
 =-2e^4(2A+B)\Big[g^{\mu\nu}g^{\rho\sigma}+g^{\mu\rho}g^{\nu\sigma}
 +g^{\mu\sigma}g^{\nu\rho}\Big]\,I_{\rm div}\,.
\end{equation}
In the limit $d\to 4$, the factor $2A+B$ vanishes as $-2\epsilon/3$ and cancels the singular term in the divergent integral \eqn{app:Idiv}, as expected. Of course, there are four-photon divergences at order $e^6$, but they cancel with contributions from the 2PI four-loop approximation, etc.

\end{document}